\documentclass[graybox]{svmult}
\usepackage{mathptmx}       
\usepackage{helvet}         
\usepackage{courier}        

\usepackage{makeidx}         
\usepackage{graphicx}        
\usepackage{multicol}        
\usepackage[bottom]{footmisc}
\usepackage{amsmath,amssymb,amsfonts}        

\makeindex             

\begin{document}

\title*{An overview of Bagger-Witten line bundles}

\author{Eric Sharpe}
\institute{Eric Sharpe @ Department of Physics MC 0435, 850 West Campus
Drive, Virginia Tech, Blacksburg, VA 24061, USA \email{ersharpe@vt.edu}}

\maketitle

\abstract{We give a brief overview of recent progress in understanding 
Bagger-Witten line bundles, which are bundles over moduli spaces of 
two-dimensional $N=2$ SCFTs whose existence is a consequence 
of the global $U(1)_R$ symmetry
of the theories.  Our overview includes a discussion of
applications in supergravities coupled to
gauge theories, a proposal for a purely geometric interpretation,
and explicit examples over moduli spaces of Calabi-Yau manifolds.
(Contribution to the proceedings of the workshop ``The geometry of
moduli spaces in string theory'' (Matrix institute, Australia,
September 2024).)  }

\section{Introduction}

Briefly, Bagger-Witten line bundles are line bundles over moduli 
spaces\footnote{
All moduli `spaces' here will be stacks, usually Deligne-Mumford stacks,
as we shall see explicitly in examples, but for simplicity and readability
we will usually refer to them as `spaces.'
} of
two-dimensional SCFTs, whose existence is implied by symmetries of the theories,
and which are closely interrelated with supersymmetry.
They were originally discovered in \cite{Witten:1982hu}, where they resolved
a puzzle in four-dimensional $N=1$ supergravity, and were later given
worldsheet realizations in terms of two-dimensional
$N=2$ SCFTs in e.g.~\cite{Distler:1992gi,Periwal:1989mx}.

We can understand Bagger-Witten line bundles as follows.
Over a moduli
space of conformal field theories, there exist bundles whose
structure groups coincide with the symmetries of the theories.
As one crosses from one coordinate patch to another over the moduli space,
across the overlaps, theories in one patch are related to theories in the
other by global symmetries.  If one has a family of theories with global 
symmetry $G$, one is naturally led to expect a principal $G$ bundle
over the moduli space, and/or vector bundles associated to such
a principal bundle.  In the case of two-dimensional $N=2$ SCFTs,
generically the only global symmetry is the $U(1)_R$ symmetry, hence a 
moduli space of two-dimensional $N=2$ SCFTs will naturally carry
a line bundle, and as it is an R symmetry, that line bundle should somehow
be related to supersymmetry.  Bagger-Witten line bundles are of this form.

Although Bagger-Witten line bundles have often been discussed and
applied (see e.g.~\cite{Silverstein:1995re}), 
their properties have historically not been
well-understood, and until recently concrete examples were not known.

The purpose of this talk is to quickly review Bagger-Witten
line bundles, culminating in some recent developments, a geometric
definition, and
concrete examples.

\section{Appearance in supergravity}  \label{sect:orig}

In this section we briefly
review the original argument of \cite{Witten:1982hu},
in four-dimensional $N=1$ supergravity, for the
existence of a line bundle over the moduli space\footnote{
It should be noted that supergravity is only meaningful over a large open
subset of the full SCFT moduli space, namely the subset `close' to the 
weak coupling limit (the large-radius limit, in a Calabi-Yau compactification).
}, and other supergravity
implications discussed since.  

Let ${\cal M}$ 
denote the moduli space of scalar field vevs.  (In a string compactification,
${\cal M}$ is a large patch on the moduli space of SCFTs, corresponding to
the weakly-coupled large-radius regime.)
Let $K$ denote the K\"ahler potential on $M$, which defines the
kinetic terms of the supergravity action.

In K\"ahler geometry, across coordinate patches,
\begin{equation}  \label{eq:kahlertrans}
K \: \mapsto \: K + f + \overline{f},
\end{equation}
where $f$ is a holomorphic function.  This also defines a symmetry of
rigidly symmetric four-dimensional $N=1$ theories.  There, the kinetic
terms can be written in superspace as  \cite[equ'n (22.1)]{Wess:1992cp}
\begin{equation}
\int d^4 \theta \, K(\Phi,\Phi^{\dag}),
\end{equation}
and as the $d^4 \theta$ annihilates purely holomorphic and antiholomorphic
functions of chiral superfields, the rigidly supersymmetric theory is
automatically invariant under~(\ref{eq:kahlertrans}).

Supergravity theories are more complicated, and the supergravity 
action turns out not to be invariant under~(\ref{eq:kahlertrans}).
For example, the superpotential terms in four-dimensional $N=1$
supergravity are proportional to
\cite[equ'n (23.3)]{Wess:1992cp}
\begin{equation}
\exp(K) \left[ g^{i j^*} \left(D_i W \right) 
\left (D_{j}
W \right)^* - 3 |W|^2 \right],
\end{equation}
which are clearly not invariant.
However, it was noted in \cite{Witten:1982hu} that
if one combines~(\ref{eq:kahlertrans}) with an action\footnote{
The action is chiral, and so in the quantum theory, anomalies must be taken
into account.  We will focus on other aspects here.
}
on the graviton $\psi_{\mu}$ and scalar superpartners $\chi^i$,
\begin{equation}
\psi_{\mu} \: \mapsto \: 
\exp\left( - \frac{i}{2} {\rm Im}\, f \right) \, \psi_{\mu},
\: \: \:
\chi^i \: \mapsto \: 
\exp\left( + \frac{i}{2} {\rm Im}\, f \right) \, \chi^i,
\end{equation}
along with a transformation of the superpotential
\begin{equation}  \label{eq:W:trans}
W \mapsto \exp(-f) \, W,
\end{equation}
then under the combined action, the theory is invariant 
under~(\ref{eq:kahlertrans}).

Now, let us examine these across elements of an open cover $\{ U_{\alpha} \}$
of ${\cal M}$ (formally treating ${\cal M}$ as a manifold),
following \cite{Witten:1982hu}.
Let $K_{\alpha}$ denote the K\"ahler potential on coordinate patch $U_{\alpha}$,
and let $f_{\alpha \beta}$ denote the holomorphic function appearing
in the transformation~(\ref{eq:kahlertrans}) across patches
$U_{\alpha}$, $U_{\beta}$ on the intersection $U_{\alpha} \cap
U_{\beta}$.  Then, for example, on the triple intersection 
$U_{\alpha} \cap U_{\beta} \cap U_{\gamma}$, one finds
\begin{eqnarray}
f_{\alpha \beta} + \overline{f}_{\alpha \beta} \: + \:
f_{\beta \gamma} + \overline{f}_{\beta \gamma} \: + \:
f_{\gamma \alpha} + \overline{f}_{\gamma \alpha}
& = &
(K_{\beta} - K_{\alpha}) +
(K_{\gamma} - K_{\beta}) +
(K_{\alpha} - K_{\gamma}),
\nonumber \\
& = & 0,
\end{eqnarray}
hence
\begin{equation}
f_{\alpha \beta} + f_{\beta \gamma} + f_{\gamma \alpha} \: = \:
h_{\alpha \beta \gamma},
\end{equation}
which are pure imaginary.  On quadruple overlaps, it is straightforward
to demonstrate that the Cech coboundary $\delta h = 0$.

If there is no gauge symmetry, one can further examine the
transformations of the scalar superpartners $\chi^i$,
following \cite{Witten:1982hu}, to argue that on triple overlaps,
\begin{equation}
\exp \left( \frac{i}{2} {\rm Im}\, h_{\alpha \beta \gamma} \right) 
\: = \: 1,
\end{equation}
hence $h_{\alpha \beta \gamma} = 4 \pi i n_{\alpha \beta \gamma}$
for integers $n_{\alpha \beta \gamma}$.  One then argues that
$\exp(-f_{\alpha \beta}/2)$ are transition functions for a line bundle,
the ``Bagger-Witten line bundle'' ${\cal L}_{\rm BW}$,
where the $(h_{\alpha \beta \gamma})$ are a Cech representative of 
$c_1$ (which is then even).

To be clear, this only defines a line bundle up to a tensor product
with a flat line bundle.
Consider transition functions, across which the K\"ahler potential
transforms as $K \mapsto K + f + f^*$.  Suppose one defines the
transition functions to be $\exp(-f/2)$, ala Bagger-witten.
Now, $K$ is invariant if one replaces $f$ by $f + g$ for $g$ pure
imaginary (hence constant, as follows from holomorphy).
After all, $g + g^* = 0$.  Hence, we could replace the transition functions
$\exp(-f/2)$ by $\exp( - (f+g)/2)$, which in general will define a different
line bundle, differing by a flat line bundle.
Hence, the supergravity
analysis above does not uniquely define a line bundle.
However, given a Bagger-Witten line bundle, the analysis
above describes how it ties into four-dimensional $N=1$ supergravity.

From the transformation law~(\ref{eq:W:trans}), we also see that the
spacetime superpotential $W$ is a (meromorphic) section
of ${\cal L}_{\rm BW}^{\otimes 2}$.
Similarly, positivity of the kinetic terms of the scalars and
their superpartners, which couple to the pullback of the metric,
was used in \cite{Witten:1982hu} to argue that ${\cal L}_{\rm BW}^{-1}$
must be ample.

More recently, the Bagger-Witten line bundle has been studied in the
context of four-dimensional gauge theories in supergravity.  
To define the gauge theory, in supergravity,
one must specify an action of the gauge
group on the scalars and their superpartners -- hence, one must specify
an action on the moduli space.  Now, a group action on a space does not
necessarily lift to a line bundle over the same space, and if it does lift,
the action will not be unique in general.
It was observed in \cite{Distler:2010zg,Hellerman:2010fv} 
that in the present case, in a gauge
theory, if the action of the gauge group on the moduli space does not
lift to the Bagger-Witten line bundle, then the action is not
(classically) invariant, and hence the gauge theory is well-defined.
Furthermore, if a lift does exist, the choice of lift is encoded
physically in the Fayet-Iliopoulos parameter, which as a result,
must be quantized (as choices of lifts are also quantized).

We can see this more explicitly, following 
\cite{Distler:2010zg,Hellerman:2010fv}.
Under a gauge transformtion, the supergravity scalars $\phi$ transform as
\begin{equation}
\delta \phi^i \: = \: \epsilon^{(a)} X^{(a) i},
\end{equation}
where $\epsilon^{(a)}$ are gauge transformation parameters and
$X^{(a) i}$ the components of a set of holomorphic Killing vectors,
the gauge field $A_{\mu}^a$ transforms as
\begin{equation}
\delta A_{\mu}^a \: = \: \partial_{\mu} \epsilon^{(a)} + f^{abc} \epsilon^{(b)}
A_{\mu}^c
\end{equation}
for $f^{abc}$ the structure constants of the Lie algebra, and
the K\"ahler potential transforms as
\begin{equation}
\delta K \: = \: \epsilon^{(a)} F^{(a)} + \epsilon^{(a)} \overline{F}^{(a)},
\end{equation}
where
\begin{equation} \label{eq:F-defn}
F^{(a)} \: = \: X^{(a)} K + i D^{(a)}.
\end{equation}
These K\"ahler potential transformations $F^{(a)}$ also appear in the
gauge transformations of the fermions, as for example
\begin{eqnarray}
\delta \chi^i & = & \epsilon^{(a)} \left( 
\frac{\partial X^{(a) i} }{\partial \phi^j} \chi^j \: + \:
\frac{i}{2} {\rm Im}\, F^{(a)} \chi^i \right),
\\
\delta \psi_{\mu} & = & - \frac{i}{2} \epsilon^{(a)} {\rm Im}\, F^{(a)} \psi_{\mu},
\end{eqnarray}
reflecting the fact that if the K\"ahler potential transforms, then the
spinors must also pick up a corresponding phase in order for the theory to
remain invariant.  This means that the infinitesimal lift of the group action
on the moduli space to the Bagger-Witten line bundle is encoded by
\begin{equation}
\frac{i}{2} \epsilon^{(a)} {\rm Im}\, F^{(a)}.
\end{equation}
Physically, from~(\ref{eq:F-defn}), we see that
shifts in the imaginary parts of
$F^{(a)}$ are the Fayet-Iliopoulos parameters, so we see that
the Fayet-Iliopoulos parameters encode a choice of lift of the
group action to the Bagger-Witten line bundle.

Let
\begin{equation}
g \: = \: \exp\left( i \epsilon^{(a)} T^a \right)
\end{equation}
be an element of the Lie group acting on the scalars, where $T^a$ is
a Lie algebra generator, then from the discussion above, the lift of $g$
to a Bagger-Witten line bundle can be described as
\begin{equation}
\tilde{g} \: = \: \exp\left( \frac{i}{2} \epsilon^{(a)} {\rm Im}\, F^{(a)}
\right).
\end{equation}
We can modify the lift by changing $\tilde{g} \mapsto \tilde{g} \exp(i \theta_g)$ 
for some collection of phases, subject to the condition that the lift
represents the group honestly:  
\begin{equation} \label{eq:honest-lift}
\tilde{g} \tilde{h} \: = \: \widetilde{gh}
\end{equation}
for all $g, h \in G$, the gauge group.
Such lifts might not always exist -- it is not guaranteed that
a set $\{ \tilde{g} \}$ exist which satisfy~(\ref{eq:honest-lift}) 
even after adding phases.  (More formally, if $\tilde{G}$ is the group
generated by the $\tilde{g}$, then a priori it is merely an extension of
$G$ by $U(1)$, and we need that extension to split in order to be able
to satisfy~(\ref{eq:honest-lift}).)

Furthermore, if a lift exists satisfying~(\ref{eq:honest-lift}),
then there are multiple lifts, obtained by deforming by phases 
$\exp(i \theta_g)$ satisfying
\begin{equation}
\exp(i \theta_g) \, \exp(i \theta_h) \: = \: \exp(i \theta_{gh} ),
\end{equation}
for all $g, h \in G$.  A collection of such $\theta_g$
defines a group homomorphism $G \rightarrow U(1)$, and so we see that if
lifts exist, they are classified by Hom$(G,U(1))$.
This means the possible lifts are quantized:
for example, if $G = U(1)$, then Hom$(G,U(1)) = {\mathbb Z}$.

We have already seen that the choice of lift of group action 
to the Bagger-Witten
line bundle is encoded by the Fayet-Iliopoulos parameters.  
Since those lifts are quantized, we see that the 
Fayet-Iliopoulos parameters are quantized.

In passing, it was noted in 
\cite{Distler:2010zg} that this story is closely related to the
geometric invariant theory (GIT) description of quotients.
There, the mathematical analogue of the Fayet-Iliopoulos parameters are also
encoded by a choice of group action on a line bundle, which (partially)
defines the GIT quotient, in the same way that a choice of point in
${\mathfrak g}^*$ partially determines a symplectic reduction.

Finally, before going on, in cases such as these, when one gauges a group
action, the moduli space is best understood as a stack, rather than a space,
following e.g.~\cite{Hellerman:2010fv,Pantev:2005wj}.
Revisiting other details of the argument of \cite{Witten:1982hu},
in \cite{Hellerman:2010fv} it was also pointed out that the transition
functions for the bundles in question only close on triple overlaps up
to a gauge transformation.  The resulting quantities can still be understood
as bundles on the corresponding stack, but are not equivalent
to ordinary bundles
on a space.

\section{SCFT description}

Bagger-Witten line bundles can also be understood more directly
over SCFT moduli spaces, see for example 
\cite{Distler:1992gi,Periwal:1989mx}.
Consider a family of conformal field theories with some global symmetry
$G$.  As one moves across coordinate patches on that family,
on overlaps, conformal field theories in one patch are related to those
on the overlapping patch by an action of $G$.  This overlap data defines
transition functions for a $G$ bundle over the parameter space.

In the case of moduli spaces of two-dimensional $N=2$ SCFTs,
since there is always at least a global $U(1)_R$ symmetry
(which exists as part of the $N=2$ superconformal algebra),
the argument above implies that there exists a line bundle over
the moduli space.  

We can then 
read off how various CFT operators transform across
the moduli space from their $U(1)_R$ charges. 
Those operators then transform globally as sections of line bundles
associated to an underlying principal $U(1)$ bundle by the
representations defined by $U(1)_R$ charges.

Two particularly important examples 
(in the notation of \cite{Lerche:1989uy})
are as follows:
\begin{itemize}
\item The spectral operator ${\cal U}_1$.  This has the same
$U(1)_R$ charge as a holomorphic top-form, and we identify the
corresponding line bundle over the moduli space with the Hodge line
bundle ${\cal L}_{\rm H}$ of holomorphic top-forms.
\item The spectral flow operator ${\cal U}_{1/2}$.  This has half the
$U(1)_R$ charge of ${\cal U}_1$, and indeed, $({\cal U}_{1/2})^2 = {\cal U}_1$.
The corresponding line bundle over the moduli space 
corresponds to a Bagger-Witten line bundle
${\cal L}_{\rm BW}$.
\end{itemize}

The SCFT description illuminates an important relationship.
Since  $({\cal U}_{1/2})^2 = {\cal U}_1$,
it is natural
to expect that Hodge line bundles and Bagger-Witten line
bundles are related by
\begin{equation}
{\cal L}_{\rm BW}^{\otimes 2} \: = \: {\cal L}_{\rm H}.
\end{equation}
Indeed, this turns out to be the case.

\section{Geometric description}

Over moduli spaces of Calabi-Yau's, a proposal for a purely
geometric description of Bagger-Witten line bundles was presented
in \cite{Donagi:2019jic,Donagi:2017mhd}.
Briefly, it was argued that 
Bagger-Witten line bundles can be interpreted as bundles of covariantly
constant spinors, in the same way that the Hodge line bundle is the
bundle of holomorphic top-forms, essentially because of their association
with the spectral flow operator ${\cal U}_{1/2}$.

On Calabi-Yau threefolds, however, there is a potential issue.
A Calabi-Yau threefold has two nowhere-zero covariantly constant spinors.
Given the description above, one is led to ask
how it can be consistent with the
original Bagger-Witten description in four-dimensional $N=1$
supergravity, as only one bundle was discussed
there.

To resolve this puzzle, we utilize the fact that the Hodge and
Bagger-Witten line bundles over moduli spaces of Calabi-Yau's are
believed to be flat.
Flatness of the Hodge line bundle over moduli spaces of Calabi-Yau's
has been discussed in e.g.~\cite{todorov2},
\cite[theorem 42, corollary 53]{todorov1},
and over moduli spaces of two-dimensional SCFTs in\footnote{
In fact, the reference \cite{Gomis:2015yaa} claimed triviality,
but as discussed in \cite{Gu:2016mxp}, their arguments only really imply
flatness, and indeed, \cite{Gu:2016mxp} gives examples which are flat but
nontrivial. See also \cite{Donagi:2017vwh} for comments on higher-dimensional cases.
}
\cite{Gomis:2015yaa}.
Since Bagger-Witten line bundles are square roots of Hodge line bundles,
their flatness follows from that of the Hodge line bundle.

Now, we return to the puzzle of consistency of the four-dimensional
$N=1$ supergravity description of Bagger-Witten line bundles
with the idea that a Calabi-Yau threefold will host two over its
moduli space.
As we noted earlier in section~\ref{sect:orig},
the original Bagger-Witten paper \cite{Witten:1982hu} only defines the
Bagger-Witten line bundle up to tensoring with a flat bundle.
Given that 
the Bagger-Witten line bundle itself is flat, the four-dimensional
$N=1$ supergravity description of \cite{Witten:1982hu} is completely
ambiguous.  In particular, in this case one can (and will) have
multiple Bagger-Witten line bundles, all of which are consistent
with the four-dimensional $N=1$ supergravity description in
\cite{Witten:1982hu}.

\section{Example: elliptic curves}

Let us begin with elliptic curves.  The moduli space
of elliptic curves, as frequently advertised, is
\begin{equation}
{\cal M}_0 \: = \: [ {\mathfrak h} / PSL(2,{\mathbb Z})],
\end{equation}
where $PSL(2,{\mathbb Z}) = SL(2,{\mathbb Z})/{\mathbb Z}_2$,
and ${\mathfrak h}$ 
denotes the upper half plane.  This space is the space of $\tau$
parameters, which transform as
\begin{equation} \label{eq:tautrans}
\tau \: \mapsto \: \tau' = \frac{a \tau + b}{c \tau +d},
\end{equation}
for 
\begin{equation}
\left[ \begin{array}{cc}
a & b \\ c & d \end{array} \right] \: \in \: SL(2,{\mathbb Z}).
\end{equation}
The ${\mathbb Z}_2$ center of $SL(2,{\mathbb Z})$, namely diagonal matrices
diag$(\pm 1, \pm 1)$, act trivially, and so they are modded out in the
description above.

Now, the Hodge line bundle is not defined over ${\cal M}_0$.
The reason for this is as follows. 
If we let $z$ denote a local coordinate on $T^2$, then under $SL(2,{\mathbb Z})$,
at the same time that $\tau$ transforms
\cite[section 2.3]{hain}
\begin{equation} \label{eq:ztrans}
z \: \mapsto \: \frac{z}{c \tau + d}.
\end{equation}
As consistency checks, let us consider a pair of special values of
$z$, namely $1$ and $\tau$.  In principle, they should map to
themselves.  It is straightforward to compute that
under the transformation above,
\begin{eqnarray}
1 & \mapsto & \frac{1}{c\tau+d} \: = \:
1 - c \left( \frac{a \tau+b}{c\tau+d} \right) + (a-1) \: \sim \: 1,
\\
\tau & \mapsto & \frac{\tau}{c\tau+d} \: = \:
\frac{a\tau+b}{c\tau+d} + (d-1) \left( \frac{a \tau+b}{c\tau+d}
\right) - b 
\: \sim \: \frac{a \tau+b}{c\tau+d},
\end{eqnarray}
where $\sim$ denotes the equivalence $z \sim z + m \tau' + n$ for
$m, n \in {\mathbb Z}$.

Returning to the actions~(\ref{eq:tautrans}), (\ref{eq:ztrans}),
we see that
although $\tau$ is invariant under the action of the center $\pm I$,
$z$ is not, and in particular, under $\pm I$, $dz \mapsto \pm dz$.

Thus, we see that holomorphic top-forms on $T^2$ are not invariant 
under the action of the center of $SL(2,{\mathbb Z})$.

As a result, the Hodge line bundle is defined over a different space,
namely
\begin{equation}
{\cal M}_1 \: = \: [ {\mathfrak h} / SL(2,{\mathbb Z})].
\end{equation}
The brackets $[]$ indicate that one takes the quotient as a stack,
and stacks distinguish quotients by trivial group actions,
so even though the difference is merely a trivially-acting group,
we see ${\cal M}_1 \neq {\cal M}_0$.

(As an aside, in physics also, gauging a trivially-acting group  
results in a different theory than 
not gauging at all.  This was discussed at length in two-dimensional
orbifolds and gauge theories in 
\cite{Pantev:2005rh,Pantev:2005zs,Pantev:2005wj}, and formed a key component of
making sense of string propagation on stacks.
In more modern language, a gauge theory in which a subgroup of the gauge
group acts trivially has a one-form symmetry, not possessed by the theory
in which that same subgroup is not gauged.)

Mathematically, ${\cal M}_1$ is a ${\mathbb Z}_2$ gerbe over
${\cal M}_0$.  The Hodge line bundle of ${\cal M}_1$ is
the generator of Pic$({\cal M}_1) = {\mathbb Z}_{12}$.

To define either Bagger-Witten bundle, we must work harder.
From the theory of orbifolds, in a supersymmetric sigma model,
given a non-R ${\mathbb Z}_2$ that maps $z \mapsto -z$,
its superpartner $\psi \mapsto - \psi$, implying that the Ramond sector vacua
map as
\begin{equation}
| \pm \rangle \: \mapsto \: \exp(\pm i \pi/2) \, | \pm \rangle \: = \:
\pm i \, | \pm \rangle.
\end{equation}
Because of the factors of $i$, more precisely because $i^2 = -1$ and not
$+1$, we see that the center of ${\mathbb Z}_2$ cannot be represented
on the Ramond vacua.  More generally, under an $SL(2,{\mathbb Z}_2)$
transformation, applying standard rules for orbifolds,
\begin{equation}
| \pm \rangle \: \mapsto \: \pm \frac{ | \pm \rangle }{ \sqrt{c \tau + d} },
\end{equation}
the same transformation as $\sqrt{dz}$.

To construct a Bagger-Witten line bundle,
we work over the moduli space
\cite{Gu:2016mxp}
\begin{equation}
{\cal M}_2 \: = \: [ {\mathfrak h} / Mp(2,{\mathbb Z}) ],
\end{equation}
where $Mp(2,{\mathbb Z})$ is the metaplectic group, the unique
nontrivial  ${\mathbb Z}_2$ central
extension of $SL(2,{\mathbb Z})$:
\begin{equation}
1 \: \longrightarrow \: {\mathbb Z}_2 \: \longrightarrow \:
Mp(2,{\mathbb Z}) \: \longrightarrow \: SL(2,{\mathbb Z})
\: \longrightarrow \: 1,
\end{equation}
whose elements can be thought of as pairs
\begin{equation}\left\{ 
\left[ \begin{array}{cc} a & b \\ c & d \end{array} \right]
\in SL(2,{\mathbb Z}), \:
\pm \sqrt{c \tau + d}
\right\},
\end{equation}
with product of the form
\begin{equation}
(A, f(-)) \cdot (B, g(-)) \: = \:
(AB, f(B(-)) g(-) ),
\end{equation}
for $A, B \in SL(2,{\mathbb Z})$.

The moduli space ${\cal M}_2$ is a ${\mathbb Z}_2$ gerbe over
${\cal M}_1$, just as ${\cal M}_1$ was itself a ${\mathbb Z}_2$ gerbe
over ${\cal M}_0$.

The Picard group Pic$({\cal M}_2) = {\mathbb Z}_{24}$, whose generator $g$
and its inverse $g^{-1}$
can be interpreted as Bagger-Witten line bundles.

This may seem like a rather abstract result, but it has concrete
implications.  For example, implicitly in the result above is an
extension of T-duality for supersymmetric sigma models on $T^2$,
from $SL(2,{\mathbb Z})$ to $Mp(2,{\mathbb Z})$.
More generally, in string duality groups, $SL(2,{\mathbb Z})$ is typically
extended to $Mp(2,{\mathbb Z})$, after one takes into account sign flips
on fermions.  This is explored in more detail in \cite{Pantev:2016nze}
(see also \cite{Tachikawa:2018njr}).

\section{Examples:  Calabi-Yau threefolds}

Moduli spaces of smooth Calabi-Yau threefolds, constructed from
orbifolds of tori, are discussed in
\cite{Donagi:2019jic,Donagi:2017mhd}.
Specifically, these papers studied moduli spaces of toroidal orbifolds
by products of ${\mathbb Z}_2$'s, classified in
\cite{Donagi:2008xy}, focusing especially on cases in which the orbifolds
have $h^{2,1} = 3$.  

At some level, the idea of the construction is to utilize results for
elliptic curves, in orbifolds of products of the form $E_1 \times E_2 \times
E_3$, where each of the $E_i$ is an elliptic curve.
For a $G$ orbifold, the moduli spaces are of the form $[ {\mathfrak h}^3 / H']$,
where $H' = H/G$ for $H$ the normalizer of the image of the orbifold
group $G$ in the maximal automorphism group of $E_1 \times E_2 \times E_3$
(which includes both
$SL(2,{\mathbb Z})^3$ and $S_3$ from exchanging the three factors).

In most of the cases with $h^{2,1} = 3$, the Picard group contains elements
of infinite order; however, in all cases, the Hodge line bundle is of
finite order.  This is a strong consistency check on
the general claim that the
Hodge line bundle should always be flat -- as the Picard group contains
elements of infinite order, not all line bundles are flat, unlike the
example of elliptic curves.
Also, as a consequence of the finite-order property, 
a globally-defined K\"ahler potential exists
over the moduli space, which is of the form
\begin{equation}
\log | \eta(\tau_1) \eta(\tau_2) \eta(\tau_3) |^2,
\end{equation}
where the $\tau_i$ are modular parameters associated with the three
elliptic curve factors.

Finally, as described in \cite{Donagi:2019jic}, 
Bagger-Witten bundles exist, and are defined on a
${\mathbb Z}_2$ gerbe over the moduli space which hosts the Hodge line bundle,
much as in the case of elliptic curves.

\section{Conclusions}

We have briefly reviewed progress in understanding Bagger-Witten line
bundles over moduli spaces of SCFTs, including recent developments
such as applications in supergravities coupled to gauge theories,
a proposal for a purely geometric interpretation, and the construction
of explicit examples over moduli spaces of Calabi-Yau manifolds.

\section{Acknowledgements}

We would like to thank J.~Distler, R.~Donagi, W.~Gu,
S.~Hellerman,
and M.~Macerato for collaborations on the topics presented here,
and also S.~Katz, T.~Pantev, and J.~Tu for 
useful discussions.
E.S. was partially supported by NSF grant PHY-2310588.

\bibliography{talk}{}
\bibliographystyle{spmpsci}

\end{document}